\PassOptionsToPackage{unicode}{hyperref}
\PassOptionsToPackage{hyphens}{url}
\documentclass[
  conf]{new-aiaa}
\usepackage{iftex}
\ifPDFTeX
  \usepackage[T1]{fontenc}
  \usepackage[utf8]{inputenc}
  \usepackage{textcomp} % provide euro and other symbols
\else % if luatex or xetex
  \usepackage{unicode-math}
  \defaultfontfeatures{Scale=MatchLowercase}
  \defaultfontfeatures[\rmfamily]{Ligatures=TeX,Scale=1}
\fi
\IfFileExists{upquote.sty}{\usepackage{upquote}}{}
\IfFileExists{microtype.sty}{% use microtype if available
  \usepackage[]{microtype}
  \UseMicrotypeSet[protrusion]{basicmath} % disable protrusion for tt fonts
}{}
\makeatletter
\@ifundefined{KOMAClassName}{% if non-KOMA class
  \IfFileExists{parskip.sty}{%
    \usepackage{parskip}
  }{% else
    \setlength{\parindent}{0pt}
    \setlength{\parskip}{6pt plus 2pt minus 1pt}}
}{% if KOMA class
  \KOMAoptions{parskip=half}}
\makeatother
\usepackage{xcolor}
\IfFileExists{xurl.sty}{\usepackage{xurl}}{} % add URL line breaks if available
\IfFileExists{bookmark.sty}{\usepackage{bookmark}}{\usepackage{hyperref}}
\hypersetup{
  pdftitle={CT-imaging in Electrostatic Thruster Ion-Optics},
  pdfauthor={Jörn Krenzer1,; Felix Reichenbach1,; Jochen Schein1,},
  hidelinks,
  pdfcreator={LaTeX via pandoc}}
\usepackage{longtable,booktabs,array}
\usepackage{calc} % for calculating minipage widths
\usepackage{etoolbox}
\makeatletter
\patchcmd\longtable{\par}{\if@noskipsec\mbox{}\fi\par}{}{}
\makeatother
\IfFileExists{footnotehyper.sty}{\usepackage{footnotehyper}}{\usepackage{footnote}}
\makesavenoteenv{longtable}
\usepackage{graphicx}
\makeatletter
\def\maxwidth{\ifdim\Gin@nat@width>\linewidth\linewidth\else\Gin@nat@width\fi}
\def\maxheight{\ifdim\Gin@nat@height>\textheight\textheight\else\Gin@nat@height\fi}
\makeatother
\setkeys{Gin}{width=\maxwidth,height=\maxheight,keepaspectratio}
\makeatletter
\def\fps@figure{htbp}
\makeatother
\providecommand{\tightlist}{%
  \setlength{\itemsep}{0pt}\setlength{\parskip}{0pt}}
\newlength{\cslhangindent}
\newlength{\csllabelwidth}
\newlength{\cslentryspacingunit} % times entry-spacing
\newenvironment{CSLReferences}[2] % #1 hanging-ident, #2 entry spacing
 {% don't indent paragraphs
  \setlength{\parindent}{0pt}
  \ifodd #1
  \let\oldpar\par
  \def\par{\hangindent=\cslhangindent\oldpar}
  \fi
  \setlength{\parskip}{#2\cslentryspacingunit}
 }%
 {}
\usepackage{calc}

\newcommand{\CSLLeftMargin}[1]{\parbox[t]{\csllabelwidth}{#1}}
\newcommand{\CSLRightInline}[1]{\parbox[t]{\linewidth - \csllabelwidth}{#1}\break}

\usepackage{iepc2022}
\IEPCsubmissionnumber{260}
\makeatletter
\@ifpackageloaded{subfig}{}{\usepackage{subfig}}
\@ifpackageloaded{caption}{}{\usepackage{caption}}
\AtBeginDocument{%

}
\AtBeginDocument{%

}
\newcounter{pandoccrossref@subfigures@footnote@counter}
{\end{figure}%
\addtocounter{footnote}{-\value{pandoccrossref@subfigures@footnote@counter}}
\@for\f:=\global@pandoccrossref@subfigures@footnotes\do{\stepcounter{footnote}\footnotetext{\f}}%
\gdef\global@pandoccrossref@subfigures@footnotes{}}
\@ifpackageloaded{float}{}{\usepackage{float}}
\floatstyle{ruled}
\@ifundefined{c@chapter}{\newfloat{codelisting}{h}{lop}}{\newfloat{codelisting}{h}{lop}[chapter]}
\floatname{codelisting}{Listing}

\makeatother
\ifLuaTeX
  \usepackage{selnolig}  % disable illegal ligatures
\fi

\title{CT-imaging in Electrostatic Thruster Ion-Optics}

\author{Jörn Krenzer\footnote{Research
Assistant, joern.krenzer@unibw.de}, Felix Reichenbach\footnote{Student
Assistant, felix.reichenbach@unibw.de}, Jochen Schein\footnote{Chairholder, jochen.schein@unibw.de}}
\affil{Laboratory for Plasma-Technology, Institute for Plasma-Technology
and Mathematics, Faculty of Electrical Engineering, Bundeswehr
University Munich, ,  ,  }

\date{2022-06-04}

\begin{document}
\maketitle

\textsuperscript{1} Laboratory for Plasma-Technology, Institute for
Plasma-Technology and Mathematics, Faculty of Electrical Engineering,
Bundeswehr University Munich

\textsuperscript{*} Correspondence:
\href{mailto:joern.krenzer@unibw.de}{Jörn Krenzer
\textless{}joern.krenzer@unibw.de\textgreater{}},
\href{mailto:felix.reichenbach@unibw.de}{Felix Reichenbach
\textless{}felix.reichenbach@unibw.de\textgreater{}},
\href{mailto:jochen.schein@unibw.de}{Jochen Schein
\textless{}jochen.schein@unibw.de\textgreater{}}

\hypertarget{introduction}{%
\section{Introduction}\label{introduction}}

The ion-optic grid-system is the essential part of electrostatic ion
thrusters governing performance and lifetime. Therefore reliable
measurements of the grid and aperture geometry over the lifetime are
necessary to understand and predict the behavior of the system. Many
different methods of measurement were introduced over the years to
tackle the challenges encountered when diagnosing single electrodes or
the whole assembly at once.

Modern industrial X-ray micro-computer-tomographs (µCT) offer the
possibility to obtain a three-dimensional density map of a grid-system
or it's components down to microscopic scales of precision. This
information allows a spectrum of new diagnostic opportunities, like
complete verification of the manufactured parts against CAD models,
detecting internal defects or density-changes or the inspection of the
assembled ion-optics and its internal alignment, which is normally
prohibited by the lack of optical access to all parts at once. Hence µCT
imaging is a promising tool to complement established methods and open
up new experimental possibilities, however it also has its own
weaknesses and pitfalls. The methods developed for grid-erosion and
-geometry measurement of a small state-of-the-art
radio-frequency-ion-thruster, the obstacles encountered along the route
will be discussed and possible solutions demonstrated.

\hypertarget{computed-tomography-fundamentals}{%
\section{Computed Tomography
Fundamentals}\label{computed-tomography-fundamentals}}

Computed tomography is a procedure where a volumetric density map is
generated from stacked images taken while orbiting in defined steps
around the specimen. A single image pixel represents the cumulated
attenuation or emission along the line-of-view to the corresponding bin
of the detector. From the stack of 2D images, a three dimensional
mapping can be reconstructed using different algorithms which are mostly
based on the \textsc{Radon}-transform first described by Johann
Radon\footnote{Johann Karl August Radon (1887-12-16 -- 1956-05-25),
  Austrian mathematician, president of the Austrian Mathematical Society}.

Even though there are many different types of tomographic imaging
techniques, the abbreviation CT is commonly understood as meaning X-ray
computed tomography due to the contemporary common usage of this imaging
procedure in health care. In this paper we will stick to this common
convention.

In the case of X-ray computed tomography (CT) the detector is a
flat-panel X-ray detector and the image taken represents the attenuation
of the radiation traversing the specimen along the line between
radiation source and the single detector-pixels. As X-ray attenuation
depends largely on the atomic mass and microscopic density of materials,
the sensor acquires a kind of integral value map of the
specimen-density.

In medical devices, the patient is kept mostly stationary while the
detector and source setup orbits around the area of interest.
Additionally the power of the X-ray source is quite high (usually in the
range of several kilowatts) while the time of exposure is kept as short
as possible, to reduce the duration of the imaging process for patient
comfort and to lessen movement-artifacts. As the risk of ionizing
radiation is well-known today, overall radiation exposure is kept to a
minimum, which constrains the amount of obtainable data and the spatial
resolution of the scans.

This is not the case for CT imaging-systems for scientific and technical
purposes, where no comfort, movement or radiation exposure issues arise.

\hypertarget{basic-uxb5ct-operation}{%
\subsection{Basic µCT Operation}\label{basic-uxb5ct-operation}}

The class of imaging systems in the field of X-ray micro-tomography
(µCT) is able to achieve spatial resolution on the scale of microns
while being able to use low-power emitters. By rotating the specimen and
keeping the detector and source stationary, the design can be
simplified. These benefits and reduced radiation hazard are paid for
with an upscaling of the scan time to several hours.

\begin{figure}
\hypertarget{fig:microct-function}{%
\centering
\includegraphics{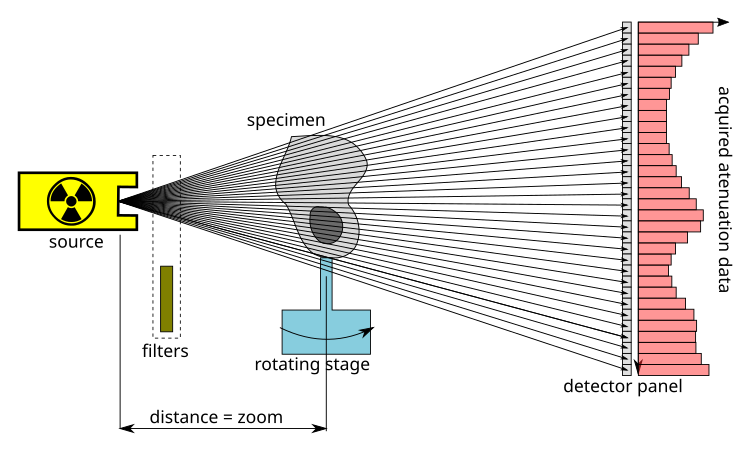}
\caption{Simplified diagram of µCT
operation}\label{fig:microct-function}
}
\end{figure}

The general mode of operation for µCT devices is illustrated in
fig.~\ref{fig:microct-function}. The X-rays emitted by the source on the
left traverse the specimen and hit the 2D flat panel detector on the
right side. The distance between source and specimen govern the
magnification factor or, in terms used by the µCT operation software,
the dimension of each pixel. As the object rotates, the minimum distance
is given by a margin necessary to prevent collision of the object and
the source. Thus, the maximum of magnification is not only constraint by
the desired area of observation but also by the geometry of the
specimen.

The exposure of each pixel corresponds to the amount of radiation
passing, which depends on thickness and attenuation of the traversed
specimen material. Simple detectors do not discriminate different
energies of the incoming X-rays, while more expensive equipment does and
thus is able to produce multi-energy stacks in one pass. To influence
the beam energy-distribution, metal filter plates can be brought in
front of the source. After each exposure a gray-scale image is saved and
the specimen is turned by a precise angle before acquisition is
repeated.

\hypertarget{sec:ct-reconstruction}{%
\subsection{Reconstruction}\label{sec:ct-reconstruction}}

Given an angle \(\theta\) around the specimen in which the image is
taken at radius \(r\), the observed attenuation
\(A_\theta \left( r \right)\) is given by
{[}\protect\hyperlink{ref-hermanFundamentalsComputerizedTomography2009}{1},
ch.~15{]}:

\begin{equation}\protect\hypertarget{eq:attenuation}{}{ A_\theta \left( r \right) = ln \left( \frac{I}{I_0} \right) = - \int \mu
\left( x, y \right) ds }\label{eq:attenuation}\end{equation}

Here, the line coordinate \(s\) signifies the orthogonal line to the
line-of-view in the origin and \(\mu\) is the location dependent
attenuation-coefficent while \(x\) and \(y\) are spatial coordinates in
the plane in which both lines lie. \(I\) is the intensity of radiation
measured, \(I_0\) is the original intensity. Using the
\textsc{Hilbert}\footnote{David Hilbert (1862-01-23 -- 1943-02-14)
  German mathematician} and the inverse
\textsc{Fourier}\footnote{Jean-Baptiste Joseph Fourier (1768-03-21 --
  1830-05-16) French mathematician and physicist}-transform, we can
solve for \(\mu\):

\begin{equation}\protect\hypertarget{eq:inverse-radon-transform}{}{ \mu \left( x, y\right) = \frac{1}{2 \pi} \int_0^\pi B_\theta \left( x \cos
\theta \cdot y \sin \theta \right) d\theta }\label{eq:inverse-radon-transform}\end{equation}

where \(B_\theta \left( x \cos \theta \cdot y \sin \theta \right)\) is
the \textsc{Hilbert}-transformed \(A_\theta \left( r \right)\). Thus, we
can reconstruct the original information using an infinite amount of
infinitesimal slices.

In practice, the amount of slices taken is limited as well as the
beam-source is not a point and the slices have a thickness. Consequently
errors are introduced which the reconstruction is indeed sensitive to
and the results thereof are artifacts and a finite resolution.
Additional inaccuracies are introduced by scattering, beam hardening,
photon statistics, flaws of the detectors used, etc. Depending on the
algorithm used, the impact of these on the reconstruction result vary.

\hypertarget{artifacts}{%
\subsection{Artifacts}\label{artifacts}}

In this section, artifacts and imaging errors will be discussed,
emphasizing those, which have major relevance for ion-optics and
thruster inspection.

\hypertarget{ring-artifacts}{%
\subsubsection{Ring Artifacts}\label{ring-artifacts}}

Failing or miscalibrated pixel units in the detector panel generate
artifactual structures akin to circular grooves. These ring artifacts
are visible in multiple slices, often concentrating around the
rotational center of the scan and can thus be differentiated from actual
structures. Reconstruction software today has quite powerful ring
artifact reduction algorithms which can suppress this kind of
distortion. This filtering process can introduce a small amount of blur
in the reconstructed images, hence sometimes reducing or disabling the
ring artifact compensation can help in enhancing visibility in features
with low contrast.

\hypertarget{beam-hardening-and-cupping}{%
\subsubsection{Beam Hardening and
Cupping}\label{beam-hardening-and-cupping}}

When the X-ray beam traverses the specimen, it gets attenuated according
to the location dependent attenuation coefficient \(\mu(\vec r)\). Each
following part of the ray will see a decreased intensity of radiation
illuminating the volume. Additionally photons with less energy will be
attenuated stronger, which works like a continuous high-pass-filter on
the photon energy distribution of the beam. Thus the density in the
depth of the mostly homogeneous specimen tends to get overestimated,
which generates a non-physical radial gradient in density.

This gradual effect is, due to its radial structure, known as
\emph{cupping artifact}. To reduce beam hardening effects, the beam can
be pre-hardened by adding a metal filter near the source, which
suppresses photons with lower energy and increases the penetrating
portion of the resulting intensity. Modern reconstruction software makes
use of polynomial correction for the gradual hardening of the beam, so
cupping can be reduced to a minimum in most cases.

\hypertarget{streaking}{%
\subsubsection{Streaking}\label{streaking}}

While gradual hardening of the radiation in homogeneous matter can be
countered quite efficiently, sharp changes in attenuation will lead to
artifacts much more difficult to reduce. Already quite significant in
the contrast of air, tissue and bone, the matter gets much worse in
technical devices comprised of plastics, ceramics and metals. The stark
contrast in attenuation generates characteristic dark bands in
reconstruction. These so called \emph{streak artifacts} occur often
positioned between dense objects or along the long axis of high
attenuation objects.

A rough estimate is, that streaking can be reduced by increasing the
mean photon energy of the beam which can be achieved by filtering or
increasing source acceleration voltage. As the X-ray generators in µCT
devices are of low energy, their photon energies are not high enough to
solve the problem in this fashion. Metal artifact reduction by using
iterative reconstruction methods is arriving in medical CT systems, but
is currently not easily available in µCT devices. Algorithms and
techniques facilitating this approach started to re-emerge in medical
applications in the last decade
{[}\protect\hyperlink{ref-mccolloughDualMultiEnergyCT2015}{2},\protect\hyperlink{ref-beisterIterativeReconstructionMethods2012}{3}{]}.
Another approach is the combination of several scans with different
energies to enhance the attenuation calculation, which is currently
evolving in recent research
{[}\protect\hyperlink{ref-caiPreliminaryStudyMultienergy2021}{4},\protect\hyperlink{ref-yaoMultienergyComputedTomography2019}{5}{]}.

\hypertarget{experimental-approach-and-experiences}{%
\section{Experimental Approach and
Experiences}\label{experimental-approach-and-experiences}}

In this section the setup, design and maturation of the experiment is
presented.

\hypertarget{uxb5ct-setup}{%
\subsection{µCT Setup}\label{uxb5ct-setup}}

\begin{figure}
\hypertarget{fig:skyscan-1173-front-1}{%
\centering
\includegraphics[width=12cm,height=\textheight]{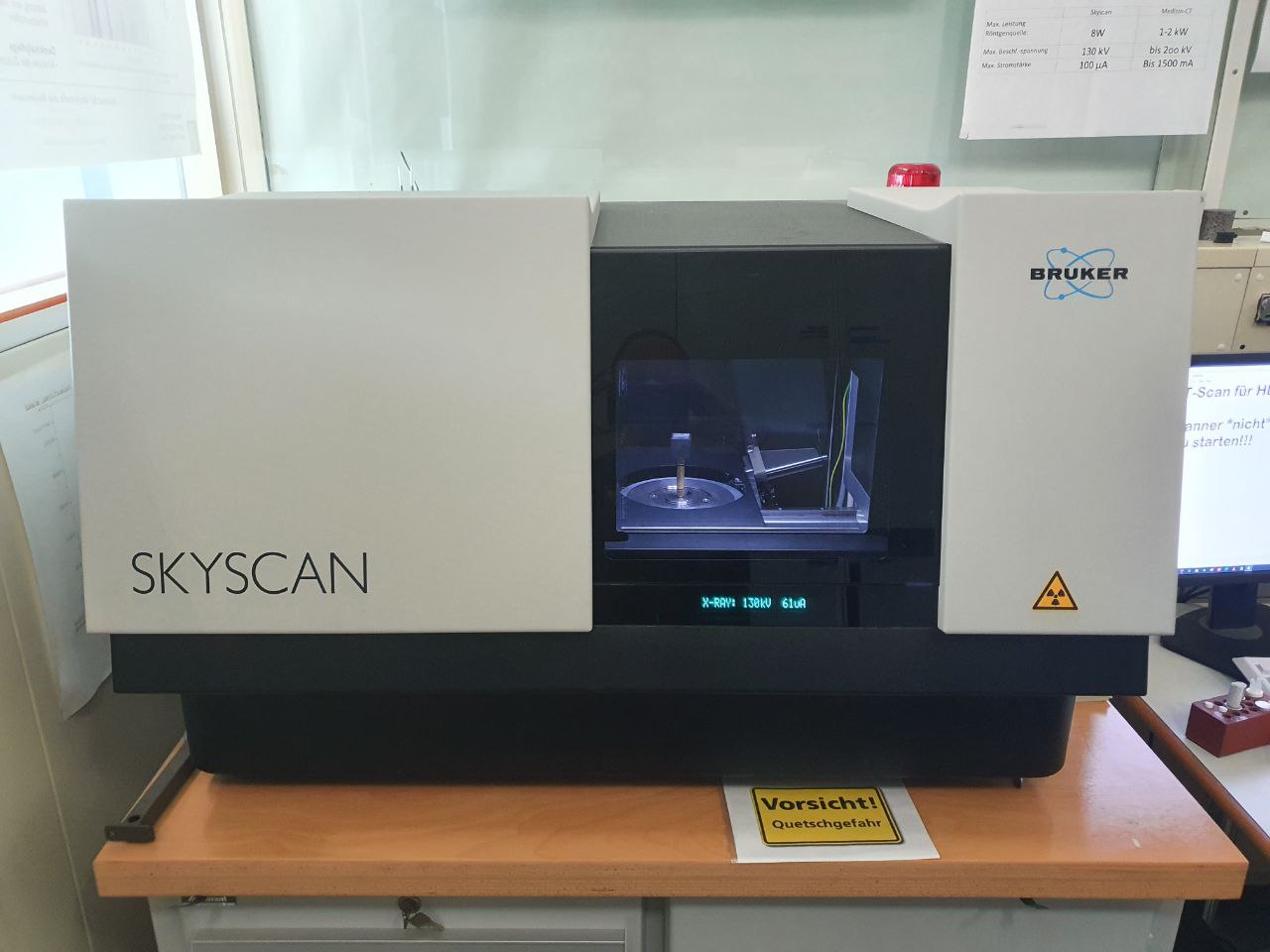}
\caption{\textsc{Bruker SkyScan 1173} µCT device front view. The lit
specimen chamber is visible above the status LED
display.}\label{fig:skyscan-1173-front-1}
}
\end{figure}

In our experiments, a commercial µCT system \textsc{SkyScan 1173} from
\textsc{Bruker}\footnote{Bruker Corporation, Billerica, MA, USA} was
used (see fig.~\ref{fig:skyscan-1173-front-1}). This model offers up to
8W of X-ray power, acceleration voltages up to 130kW and a flat panel
detector with 2240 by 2240 pixels. The source and detector setup ist
stationary, while the specimen is positioned by a movable stage to
adjust viewing location and magnification. In tomographic image
acquisition the specimen is incrementally rotated by the stage.

As the scanning specimen, a used ion-optic of a state-of-the-art
flight-hardware small electrostatic radio-frequency-ion-thruster was
utilized in assembled and disassembled state. From previous experiments
in the scope of research funded by the European Space Agency (ESA),
experience with this type of grid system was already acquired
{[}\protect\hyperlink{ref-esa400012491218NL2018}{6}{]}.

In course of this investigation, scanning of the assembled ion-optics
has revealed problematic disturbances, so a phantom for further research
into these matters has been designed and used. See
sec.~\ref{sec:phantom-design} for more information.

\hypertarget{scanning-regimes-for-ion-optic-inspection}{%
\subsection{Scanning Regimes for Ion-Optic
Inspection}\label{scanning-regimes-for-ion-optic-inspection}}

\begin{figure}
\hypertarget{fig:skyscan-1173-specimen-chamber-1}{%
\centering
\includegraphics[width=15cm,height=\textheight]{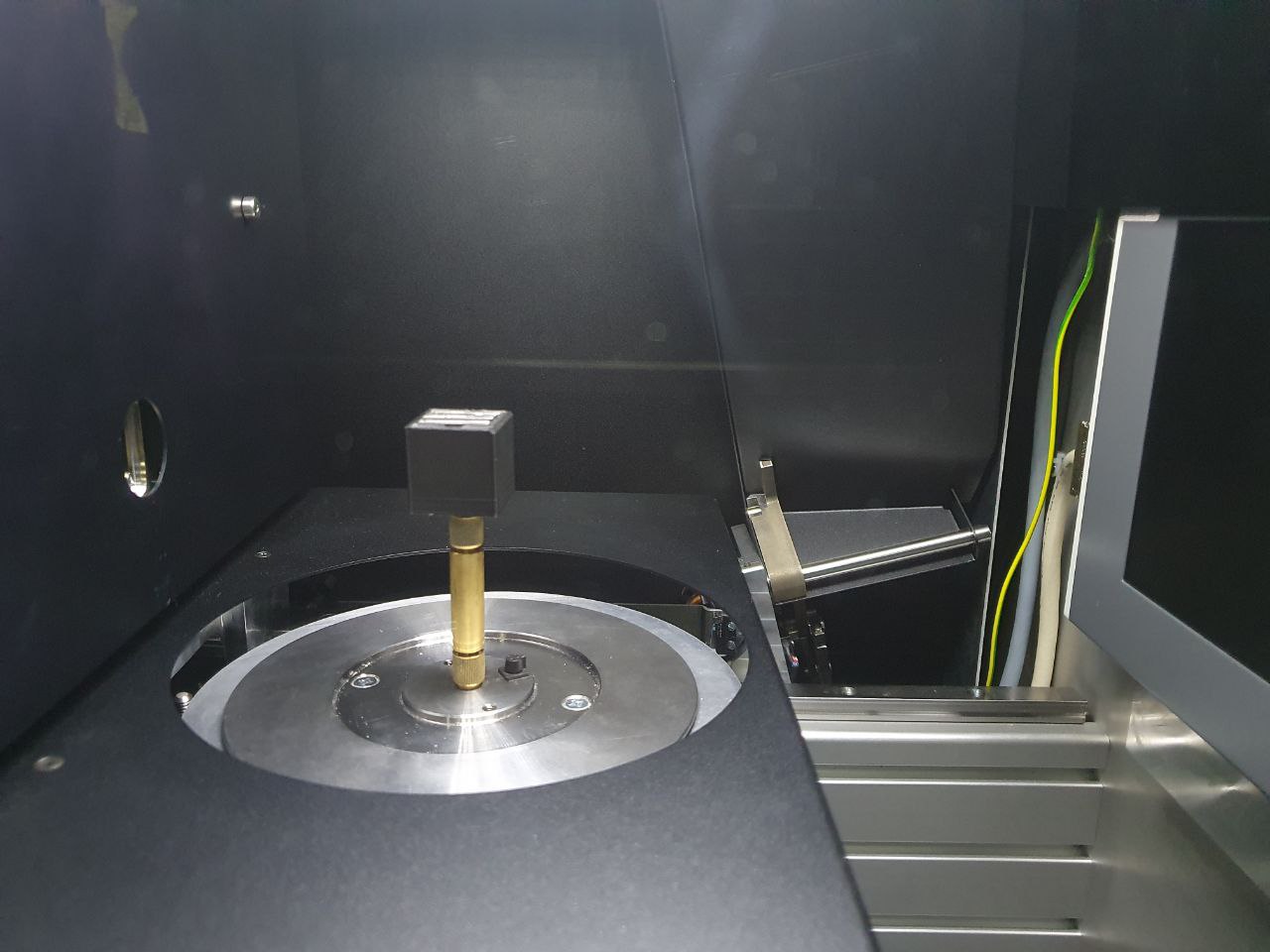}
\caption{\textsc{Bruker SkyScan 1173} specimen chamber. View from the
front into the device. In the specimen chamber covered by
radiation-absorbent glass the X-ray tube is on the left, the
specimen-stage in the middle and the detector-panel on the
right.}\label{fig:skyscan-1173-specimen-chamber-1}
}
\end{figure}

In general, specimen for µCT scanning have to be mounted securely to the
stage of the device used. Any room for internal or stage-relative
movement can introduce errors by slight positioning changes of the
examined object or it's components. The maximum magnification is
constrained by the radial bounds of the object, so it is favorable to
align the longest axis of the object with the stage's rotation axis.
Before introduction of the specimen, manufacturer procedures for warm-up
and calibration of the device have to be followed. Furthermore, before
beginning the scanning cycle clearance to all scanner parts for a full
rotation must be ensured (see
fig.~\ref{fig:skyscan-1173-specimen-chamber-1}).

If multiple scans at different X-ray settings for the same object are
desired, e.g.~for multi-energy CT applications or dataset blending, the
device must not be reset or restarted, in order to avoid positioning
differences introduced by automatic recalibration of the stage
positioning system. Unnecessary movements of the stage between the takes
should be avoided to not accumulate errors by steps missed by the
articulating stepper motors. Nevertheless, registration of the different
stacks should be checked after acquisition and, if applicable,
corrected.

When inspecting thruster ion-optics, this can be done in two different
ways:

\begin{enumerate}
\def\labelenumi{\arabic{enumi}.}
\tightlist
\item
  Disassembled, one component at a time
\item
  Assembled, the whole system at once
\end{enumerate}

Option 1 imposes less constraints and, as experience shows, less
artifacts and distortion due to the mostly uniform material. It has
proven especially valuable for in-detail inspection of component
dimension and erosion state. However, it is not possible to check
alignment, positioning and the status of all parts as a unit this way.
Additionally, each consecutive cycle of disassembling and assembling can
introduce wear, damage or changes to system alignment which can impact
performance data measure.

Option 2 offers the possibility to verify the state of the whole system,
e.g.~to document correct and unharmed assembly, fathom damage or failure
causes or to investigate alignment issues. Scanning and reconstructing
assembled grid-systems has been found difficult due to contrast issues
and especially streaking near and around the aperture region. Results so
far enable basic check of alignment and geometry using the outer
features, but no in-depth inspection of the ion-optic system in the
active region.

\hypertarget{sec:phantom-design}{%
\subsection{Phantom Design}\label{sec:phantom-design}}

Experimental experience has shown that the inspection of assembled
extraction grid systems is hampered by strong streaking artifacts
generated due to the patterned structure of very differently attenuating
materials in close proximity. On-board software of system manufacturer
and available beam pre-hardening capabilities of the device are not able
to compensate the detrimental level of artifacts.

\begin{figure}
\hypertarget{fig:phantom-overview}{%
\centering
\includegraphics{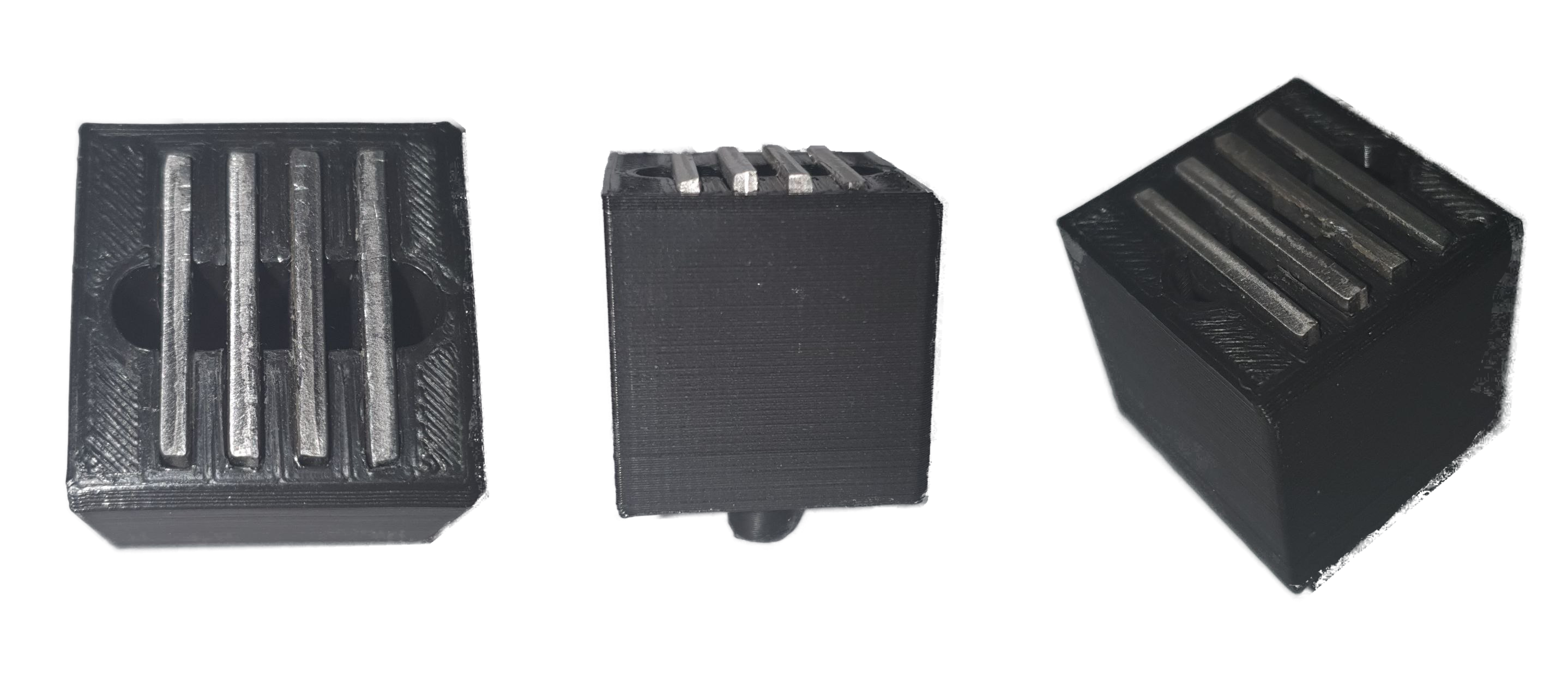}
\caption{Phantom used for research into streaking mitigation for
assembled grind system inspections. Top, left and diagonal view.
Rotation axis is parallel to the stainless steel plates and is centered
in the mounting rod (cylindrical protrusion visible on the bottom of the
left-side-view).}\label{fig:phantom-overview}
}
\end{figure}

To research strategies for countering streaking, so a full inspection of
the assembled grid system can be made possible, a phantom was designed
which reproduces the artifacts generated by the grid system (see
fig.~\ref{fig:phantom-overview}). It was found by comparison, that the
strongest observed artifacts can be generated by placing parallel
high-attenuating metal plates in close proximity into a low-attenuating
material. Four metal plates were chosen to deliver strong artifact
noise, so a solution showing promising enhancements can be expected to
perform comparable or better on a real grid system. For ease of
manufacturing, stainless steel was chosen for the metal, while the
holding matrix was 3D-printed from polylactide (PLA).

\hypertarget{reconstruction-and-postprocessing-software}{%
\subsection{Reconstruction and Postprocessing
Software}\label{reconstruction-and-postprocessing-software}}

For reconstruction and data post-processing the manufacturer's
\textsc{nRecon} software was used. Optimal parameters for each
reconstruction have to be manually selected by experience and choosing
best performing settings in multiple preview images generated at
different areas-of-interest. The software suite accompanying the device
also offers basic abilities for data visualization, analysis and export
into different formats for further processing.

\hypertarget{software-development-endeavor}{%
\subsection{Software Development
Endeavor}\label{software-development-endeavor}}

While acquisition of volumetric data for single components of the grid
systems of small ion thrusters could be achieved with satisfactory
results, inspection of assembled components proved to be difficult due
to strong streaking. In an effort to reduce disturbances by artifacts,
development of preprocessing software was started. The approach taken
was to combine images taken at different settings akin to exposure
bracketing practiced in photography, to compensate for the
overestimation of attenuation in proximity of the high attenuating
structures.

Benefits which where identified for this approach are the ability to
still use the manufacturer-provided tool chain, evasion of the high
complexity of writing a standalone solution using multi-energy-beam or
iterative-reconstruction-approaches and much less demanding hardware
requirements as the aforementioned algorithms would impose.

A command-line-interface (CLI) \textsc{python 3} software was developed,
using the libraries \textsc{OpenCV}, \textsc{SciPy} and \textsc{NumPy}
to enable generation of combined datasets from multiple input stacks
{[}\protect\hyperlink{ref-opencv_library}{7}--\protect\hyperlink{ref-2020SciPy-NMeth}{9}{]}.

\hypertarget{discussion}{%
\section{Discussion}\label{discussion}}

\hypertarget{results-of-single-grid-electrode-inspection}{%
\subsection{Results of Single Grid Electrode
Inspection}\label{results-of-single-grid-electrode-inspection}}

Imaging of single components of low to medium attenuation has provided
good results. Spatial resolution of at least \(30\mu m\) (three times
pixel-size of \(10 \mu m\)) could be achieved, constrained by the
physical dimensions of the components.

\begin{figure}
\hypertarget{fig:uct-microscope-comparison}{%
\centering
\includegraphics{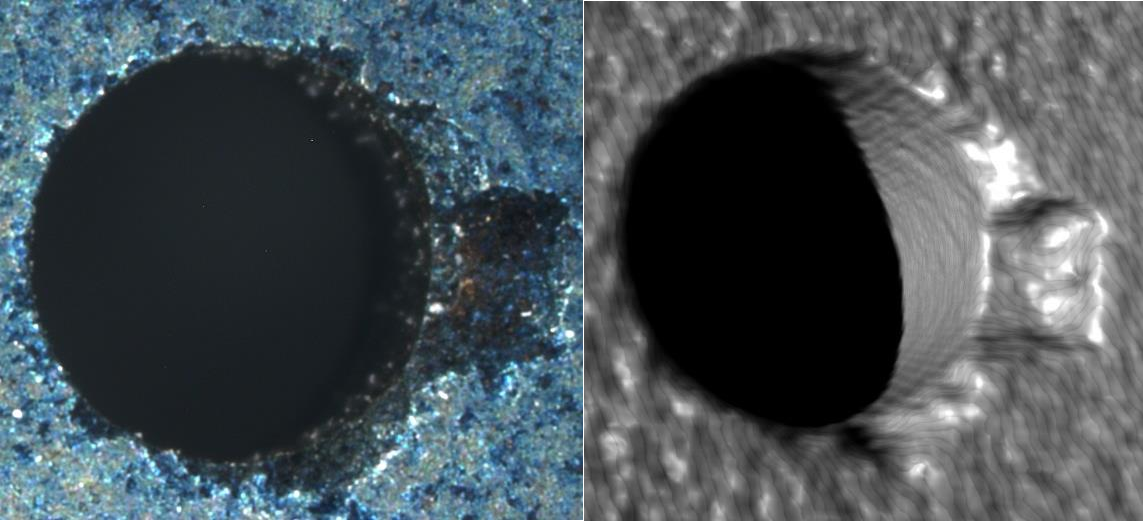}
\caption{Comparison of the reconstructed surface with a microscopy
photo. Pixel size of the reconstruction is approx. \(11 \mu m\).
Material used for electrode is
graphite.{[}\protect\hyperlink{ref-reichenbachAdvancedInspectionMeasurement2020}{10}{]}}\label{fig:uct-microscope-comparison}
}
\end{figure}

Features identified by the µCT reconstruction could be verified against
visual microscopy as demonstrated in
fig.~\ref{fig:uct-microscope-comparison}
{[}\protect\hyperlink{ref-reichenbachAdvancedInspectionMeasurement2020}{10}{]}.
Three dimensional geometry information was extracted in scope of
previous activity funded by ESA and has proven to be valuable for
quality assurance and erosion monitoring
{[}\protect\hyperlink{ref-esa400012491218NL2018}{6}{]}.

\begin{figure}
\hypertarget{fig:screen-streaking}{%
\centering
\includegraphics{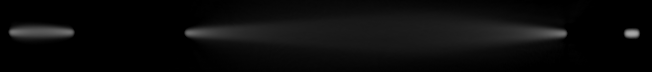}
\caption{Streaking observable in the cut-view of a thin molybdenum
electrode. The artifacts occur at the long stretches of metal while the
small structures are less influenced.}\label{fig:screen-streaking}
}
\end{figure}

\begin{figure}
\hypertarget{fig:phantom-histogram-head-on}{%
\centering
\includegraphics{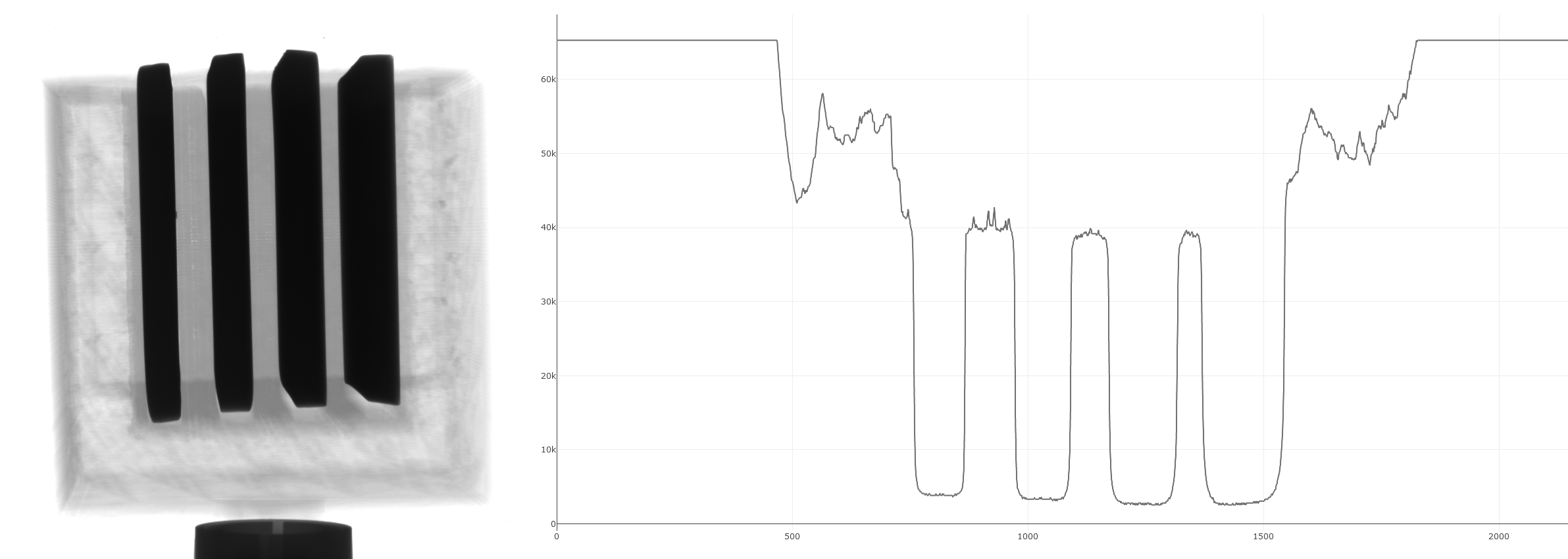}
\caption{X-ray view of the phantom along the stainless steel plates.
Even attenuated by \(20 mm\) of steel the illumination is well above
zero on a medium exposure time.}\label{fig:phantom-histogram-head-on}
}
\end{figure}

In components made of high attenuation material, like steel or
molybdenum (see fig.~\ref{fig:screen-streaking}) strong streaking and
beam hardening artifacts were present along more massive sections. As
the X-ray images taken are of good quality and show enough illumination
in all areas, see fig.~\ref{fig:phantom-histogram-head-on}, this
problems are likely to stem from the reconstruction algorithm used by
the manufacturer's software. Even though the used algorithm remains
undisclosed, regarding the sensitivity to high attenuation objects and
artifacts, probably a variant of the Filtered Back Projection (FBP)
algorithm is implemented in the tool chain
{[}\protect\hyperlink{ref-hermanFundamentalsComputerizedTomography2009}{1},
ch.~10{]}.

As FBP is know for it weaknesses and mainly used for ease of
implementation and low demand on calculation power, the problems
observed with high attenuation materials might be mitigated by using a
different reconstruction algorithm in future investigations, as X-ray
illumination seems sufficient.

\hypertarget{results-of-assembled-ion-optics-inspection}{%
\subsection{Results of Assembled Ion-Optics
Inspection}\label{results-of-assembled-ion-optics-inspection}}

As discussed before, scanning the assembled ion-optics has proven prone
to strong artifact noise in the region of interest surrounding the
active aperture area. Even though basic alignment could be checked by
using features in the outer regions, inspection for correct positioning
or damage in components near the beam area was impossible. To
investigate causes and possible solutions, a phantom was designed.

\hypertarget{results-of-investigations-with-phantom}{%
\subsection{Results of Investigations with
Phantom}\label{results-of-investigations-with-phantom}}

The phantom developed has shown strong artifacts in CT reconstruction
attempts and has clearly made the limitations and problems of the µCT
system in regard of imaging typical ion-optics arrangements visible.
Attempts to mitigate the issues by common advice such as pre-hardening
the beam, using beam hardening correction and choosing high X-ray
voltage, were not successful, even though small improvements could be
realized.

\hypertarget{own-software-approach}{%
\subsubsection{Own Software Approach}\label{own-software-approach}}

By using the CLI-program developed for this activity, improvements in
visibility in the phantoms area of interest could be achieved. It was
sufficient to run the software on at least two stacks of scans at
different energies and exposure settings and to reconstruct the result
with the manufacturer's tool chain to achieve visible improvements.

\begin{figure}
\hypertarget{fig:comparison-input-blended-1}{%
\centering
\includegraphics{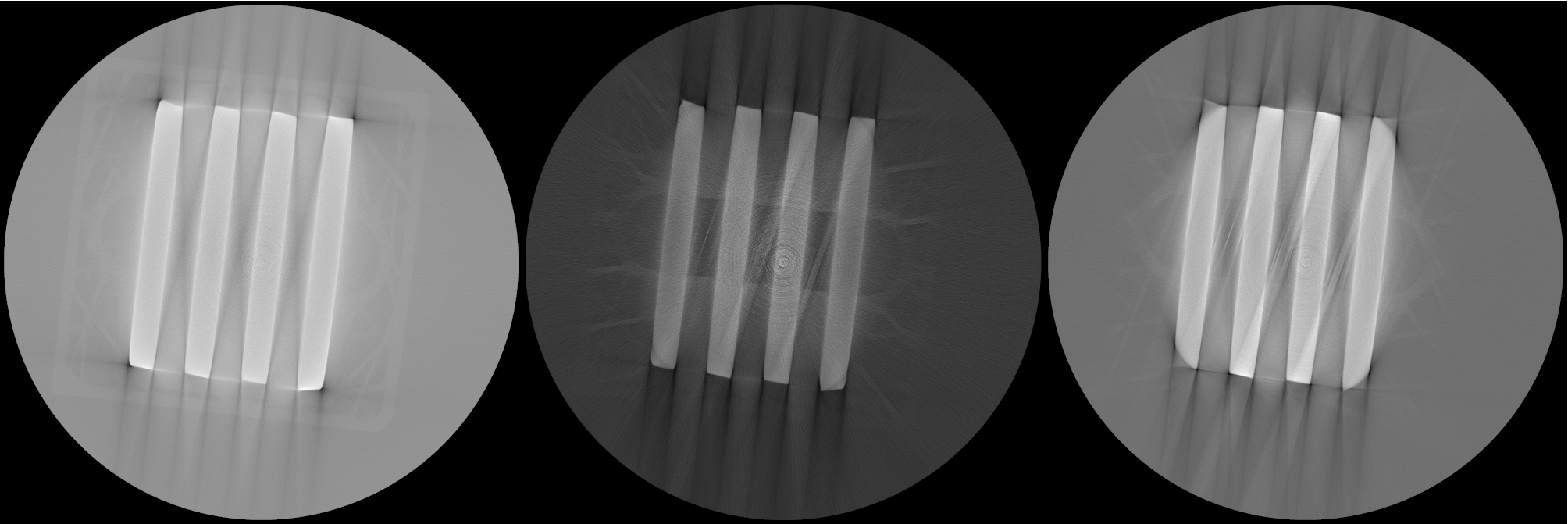}
\caption{Examples of raw reconstruction for two stacks scanned of the
phantom (left and right) and the blended stack (center) without ring
artifact and beam hardening compensation. X-ray setting was \(130kV\) at
\(61\mu A\), exposure for left stack \(1000ms\) for right stack
\(4000ms\). While outer PLA structure is well visible in left stack, low
attenuating features in the area between the plates is lost. Inner PLA
structure is visible in the right stack, but massive streaking occurs
and outer structure vanishes. Blended stack retains most of the outer
structure and shows inner features more
clearly.}\label{fig:comparison-input-blended-1}
}
\end{figure}

Best results were obtained using a simple 16bit fusing algorithm, while
unmodified high-dynamik-range (HDR) blending algorithms provided by
\textsc{OpenCV} and \textsc{SciPy} have introduced to much noise in the
resulting reconstruction. An example is shown in
fig.~\ref{fig:comparison-input-blended-1}, without artifact compensation
and at automatic determined similar contrast settings to make the
difference comparable.

As the manufacturer's reconstruction is prone to strong artifacts when
dealing with high attenuation contrasts, it has been deemed necessary to
combine the multi-energy scanning approach with more robust iterative
reconstruction algorithms to achieve even better results in the future.
For this reconstruction capabilities will have to be added to the
developed software.

Iterative reconstruction is currently re-emerging, as growing dataset
sizes in early CT development made implementation on contemporary
hardware impractical and thus these methods fell into disuse. With the
advent of modern highly parallel computing capacities, this kind of
reconstruction becomes feasible again in practical timescales
{[}\protect\hyperlink{ref-hermanFundamentalsComputerizedTomography2009}{1},\protect\hyperlink{ref-beisterIterativeReconstructionMethods2012}{3}{]}.
Even though much more demanding on computation time, these algorithms
are superior in flexibility and generate much less artifacts than the
common FBP approach
{[}\protect\hyperlink{ref-hermanFundamentalsComputerizedTomography2009}{1},
ch.~11{]}.

\hypertarget{outlook}{%
\subsection{Outlook}\label{outlook}}

Future work should strive to employ better reconstruction algorithms
under the control of the experimenter to further improve result quality
for high attenuation material and assembled ion-optic scanning. The
ability to precisely document and analyze the ion-optic system and it's
components has the potential to add great value for quality assurance
and research into erosion and failure mechanisms. Currently emerging
specialized multi-energy reconstruction algorithms might prove even more
capable in obtaining high resolution undistorted reconstructions of the
active aperture area of assembled ion-optics.

While single component scanning already delivers good results for some
materials, improvements found for assembled systems could help to
enhance results for highly attenuating materials.

Algorithms for discerning materials from multi-energy reconstructions
could be employed to gain insight into deposition and chemical processes
occurring on the ion-optics over their lifetime.

\hypertarget{conclusion}{%
\section{Conclusion}\label{conclusion}}

µCT imaging has the potential to enable full volume documentation and
analysis of thruster ion-optics and it's components. Scanning of single
components made of low to medium X-ray attenuating materials delivers
good results when using the toolchain delivered with a typical
commericially available µCT device. Even though the operator has to
determine the optimal parameters according to each object, material and
region-of-interest combination, given some experience, valuable data can
be obtained in a reasonable amount of time.

Scanning objects which attenuate the radiation stronger or whole
ion-optic assemblies is prone to strong artifacts and distortions due to
the reconstruction algorithms currently in wide spread use in µCT
systems' on-board software. To solve these obstacles, independent
reconstruction software has to be used or developed, which relies on
more modern and robust reconstruction methods. With single objects
reconstruction is sufficiently clear in the aperture area, where
material thickness is quite low. For assemblies basic alignment and
positioning can be checked in the outer areas, were there tend to be
less artifacts.

Simple blending of image stacks taken at different scanner settings can
alleviate some of the artifacts encountered when using the
manufacturer-provided tool chain on assembled ion-optics without the
need to develop or use special software. Combining multiple scans at
different energies with enhanced reconstruction algorithms, especially
those developed to make use of the additional information in the
reconstruction process itself, are deemed the way to go to develop good
quality full volumetric mapping. Additional value might come from
methods which discern specimen materials from multi-energy-scan data.

\hypertarget{acknowledgments}{%
\section*{Acknowledgments}\label{acknowledgments}}
\addcontentsline{toc}{section}{Acknowledgments}

Part of this work was kindly sponsored by the General Support Technology
Programme (GSTP) of the European Space Agency (ESA)
{[}\protect\hyperlink{ref-esa400012491218NL2018}{6}{]}. Please refer to
the references section for details.

\hypertarget{references}{%
\section*{References}\label{references}}
\addcontentsline{toc}{section}{References}

\hypertarget{refs}{}
\begin{CSLReferences}{0}{0}
\leavevmode\vadjust pre{\hypertarget{ref-hermanFundamentalsComputerizedTomography2009}{}}%
\CSLLeftMargin{{[}1{]} }%
\CSLRightInline{Herman, G. T.
\emph{\href{https://doi.org/10.1007/978-1-84628-723-7}{Fundamentals of
Computerized Tomography}}. Springer London, London, 2009.}

\leavevmode\vadjust pre{\hypertarget{ref-mccolloughDualMultiEnergyCT2015}{}}%
\CSLLeftMargin{{[}2{]} }%
\CSLRightInline{McCollough, C. H., Leng, S., Yu, L., and Fletcher, J. G.
{``Dual- and Multi-Energy CT: Principles, Technical Approaches, and
Clinical Applications.''} \emph{Radiology}, Vol. 276, No. 3, 2015, pp.
637--653. \url{https://doi.org/gg8pb6}.}

\leavevmode\vadjust pre{\hypertarget{ref-beisterIterativeReconstructionMethods2012}{}}%
\CSLLeftMargin{{[}3{]} }%
\CSLRightInline{Beister, M., Kolditz, D., and Kalender, W. A.
{``Iterative Reconstruction Methods in X-Ray CT.''} \emph{Physica
Medica}, Vol. 28, No. 2, 2012, pp. 94--108.
\url{https://doi.org/gh22w7}.}

\leavevmode\vadjust pre{\hypertarget{ref-caiPreliminaryStudyMultienergy2021}{}}%
\CSLLeftMargin{{[}4{]} }%
\CSLRightInline{Cai, A., Zhong, X., Yu, X., Wang, Y., Li, L., and Yan,
B. \href{https://doi.org/10.1109/ICMIPE53131.2021.9698896}{A Preliminary
Study on Multi-Energy CT Reconstruction via Weighted Tensor Nuclear Norm
Combining Image Sparsity}. Presented at the 2021 IEEE International
Conference on Medical Imaging Physics and Engineering (ICMIPE), Hefei,
China, 2021.}

\leavevmode\vadjust pre{\hypertarget{ref-yaoMultienergyComputedTomography2019}{}}%
\CSLLeftMargin{{[}5{]} }%
\CSLRightInline{Yao, L., Zeng, D., Chen, G., Liao, Y., Li, S., Zhang,
Y., Wang, Y., Tao, X., Niu, S., Lv, Q., Bian, Z., Ma, J., and Huang, J.
{``Multi-Energy Computed Tomography Reconstruction Using a Nonlocal
Spectral Similarity Model.''} \emph{Physics in Medicine \& Biology},
Vol. 64, No. 3, 2019, p. 035018.
\url{https://doi.org/10.1088/1361-6560/aafa99}.}

\leavevmode\vadjust pre{\hypertarget{ref-esa400012491218NL2018}{}}%
\CSLLeftMargin{{[}6{]} }%
\CSLRightInline{ESA, Ed. 4000124912/18/NL/KML -- Improvement Of The
Lifetime Of Electric Propulsion Thrusters Using Different Propellant By
Reducing Sputtering Effects On Materials., 2018.}

\leavevmode\vadjust pre{\hypertarget{ref-opencv_library}{}}%
\CSLLeftMargin{{[}7{]} }%
\CSLRightInline{Bradski, G. {``The OpenCV Library.''} \emph{Dr.~Dobb's
Journal of Software Tools}, 2000.}

\leavevmode\vadjust pre{\hypertarget{ref-harris2020array}{}}%
\CSLLeftMargin{{[}8{]} }%
\CSLRightInline{Harris, C. R., Millman, K. J., van der Walt, S. J.,
Gommers, R., Virtanen, P., Cournapeau, D., Wieser, E., Taylor, J., Berg,
S., Smith, N. J., Kern, R., Picus, M., Hoyer, S., van Kerkwijk, M. H.,
Brett, M., Haldane, A., del Río, J. F., Wiebe, M., Peterson, P.,
Gérard-Marchant, P., Sheppard, K., Reddy, T., Weckesser, W., Abbasi, H.,
Gohlke, C., and Oliphant, T. E. {``Array Programming with NumPy.''}
\emph{Nature}, Vol. 585, No. 7825, 2020, pp. 357--362.
\url{https://doi.org/ghbzf2}.}

\leavevmode\vadjust pre{\hypertarget{ref-2020SciPy-NMeth}{}}%
\CSLLeftMargin{{[}9{]} }%
\CSLRightInline{Virtanen, P., Gommers, R., Oliphant, T. E., Haberland,
M., Reddy, T., Cournapeau, D., Burovski, E., Peterson, P., Weckesser,
W., Bright, J., van der Walt, S. J., Brett, M., Wilson, J., Millman, K.
J., Mayorov, N., Nelson, A. R. J., Jones, E., Kern, R., Larson, E.,
Carey, C. J., Polat, İ., Feng, Y., Moore, E. W., VanderPlas, J.,
Laxalde, D., Perktold, J., Cimrman, R., Henriksen, I., Quintero, E. A.,
Harris, C. R., Archibald, A. M., Ribeiro, A. H., Pedregosa, F., van
Mulbregt, P., and SciPy 1.0 Contributors. {``SciPy 1.0: Fundamental
Algorithms for Scientific Computing in Python.''} \emph{Nature Methods},
Vol. 17, 2020, pp. 261--272. \url{https://doi.org/ggj45f}.}

\leavevmode\vadjust pre{\hypertarget{ref-reichenbachAdvancedInspectionMeasurement2020}{}}%
\CSLLeftMargin{{[}10{]} }%
\CSLRightInline{Reichenbach, F. \emph{Advanced Inspection and
Measurement of Grid-Geometry for Refined Predictions of Wear in
Electrostatic Ion-Thrusters}. Bachelor Thesis. Bundeswehr University
Munich, Neubiberg, 2020.}

\end{CSLReferences}

\end{document}